\documentclass[12pt]{article}
\title{\bf {Constraint On The Cosmological Constant From Gravitational Lenses In An  Evolutionary Model  Of Galaxies  }}
\author{Deepak Jain\thanks{E--mail : deepak@ducos.ernet.in},
N.Panchapakesan\thanks{E--mail : panchu@ducos.ernet.in},
S.Mahajan\thanks{E--mail : sm@ducos.ernet.in} and 
V.B.Bhatia\thanks{E--mail : vbb@ducos.ernet.in} \\
	{\em Department of Physics and Astrophysics} \\
        {\em University of Delhi, Delhi-110 007, India} 
	}

\begin {document}
\maketitle

%\begin {abstract}
%\large
%\baselineskip=15pt
%  \input {abstract.tex}
%   \end {abstract}
%   \vfill
%   \eject

%\begin {document}
%\maketitle

\begin {center}
\Large {\bf Abstract}
%\large
%\baselineskip=15pt 
%\input {abstract.tex}
\end{center}
%\begin {abstract}
\large
\baselineskip=15pt
  We study the effect of the cosmological constant on the statistical
properties of gravitational lenses in flat cosmologies
$(\Omega_{0}+\lambda_{0} = 1)$. It is shown that some of the lens
observables are strongly affected by the cosmological constant,
especially in a low--density universe, and its existence might be
inferred by a statistical study of the lenses. In particular, the
optical depth of the lens distribution may be used best for this
purpose without depending much on the lens model.
We calculate the optical depth (probabilty of a beam encountering with
a lens event) for a source in a new picture of galaxy evolution based
on number evolution in addition to pure luminosity evolution.  It seems that present
day galaxies result from the merging of a large number of building
blocks. We have tried to put limit on the cosmological constant in this new picture of galaxy evolution. This evolutionary model of galaxies permits larger value of cosmological constant.

%     \end {abstract}
      \vfill
      \eject

\large
%\baselineskip=2\baselineskip
\baselineskip = 20pt
\begin {section} {Introduction}
   In the early history of modern cosmology the cosmological constant $\Lambda$ was
invoked twice. First by Einstein to obtain static models of the
universe. Next by Bondi, Gold and Hoyle to resolve an age crisis and to
construct the universe that satisfied the `` Perfect Cosmological
Principle'', i.e., the hypothesis that the universe appears the same at all times and places. In
both instances the motivating crisis passed and the cosmological
constant was put aside. However the cosmolgical constant remains a focal point
of cosmology and of particle theory. The former because today an understanding of a  wide
range of observations seem to call for a cosmological constant. The
latter because in the context of quantum--field theory a cosmological
constant corresponds to the energy density associated with the vacuum
and no known principle demands that it vanish.

Various aspects of  recent observations  suggest
reconsideration of a nonvanishing cosmological constant (Krauss \& Turner 1995). These includes
the age of universe once again, the formation of large--scale
structure (galaxies,clusters of galaxies, superclusters) and the
matter content of the universe as constrained by dynamical
estimates, Big Bang Nucleosynthesis and X--Ray observations of
clusters of galaxies.

One of the possibilities of detecting  $\Lambda$ is the gravitational lensing technique. To use the gravitational lensing as a tool for the
determination of cosmological parameters either by a detailed study of
specific lens systems or through statistical analysis of a sample of
lenses has been  frequently discussed (Refsdal 1964; Press \& Gunn 1973). It has been
pointed out that the expected frequency of multiple imaging lensing
events is quite sensitive to  cosmological constant (Fukugita, Futamse \& Kasai 1990; Turner 1990). To put a limit
on  $\Lambda$,  we need to first calculate the expected number of multiple image
 gravitational lens systems (produced by the known galaxy population) to 
be expected in a particular quasar sample with a known distribution
of redshifts. This then has to be compared  with the observed frequency of lens systems found.

We have used a unifying model of galaxy formation which can answer 
two questions raised by the recent data on high - redshift galaxies.
 These concern the ages of, and star formation history in the distant 
radiogalaxies, and the nature of the large number of field galaxies 
revealed by faint galaxy counts. This new model is based on the strong 
number evolution in addition to pure luminosity evolution of the galaxies (Volmerange \& Guiderdoni 1990).

In Section 2, we write down the New Luminosity Function (NLF) which
has strong number evolution in addition to pure luminosity
evolution. In Section 3, we present a new calculation of the total
multiple image lensing cross--sections for a galaxy in the Singular Isothermal
Sphere (SIS) approximation (Turner, Ostriker \& Gott 1984)(hereafter TOG) based on the new galaxy luminosity
function, velocity--luminosity correlation and velocity dispersion. In
Section 4 we write down the basic equations for the statistical
properties of lenses for each of the  galaxy models. The quasar luminosity
function for the BSP sample (Boyle, Shanks \& Peterson 1988) is described in Section 5.  Section 6 contains 
the number of lensed quasars in the BSP sample for the Press--Schecter Luminosity
Function (PSLF) and the New Luminosity Function (NLF). In Section 7 we discuss the results.

   \end {section}

\begin {section} {New Luminosity Function}
   In 1990, Volmerange and Guiderdoni, proposed a unifying model to
explain faint galaxy counts as well as observational properties of
distant radiogalaxies. This new model of galaxy evolution is based on number evolution in addition to pure luminosity evolution. Present day galaxies result from the merging of a large number of building blocks and the comoving number of these building blocks evolves as $ ( 1 + z)^{1.5}$. It is argued that the present luminosity function is the well
known Press - Schecter Luminosity Function (PSLF)$$\Phi(L,z = 0) =
\phi^\ast (L /L_\ast)^\alpha exp(- L/ L_\ast) dL/L_\ast$$
with $L_\ast$ being the characterstic luminosity at the knee and
$\phi^\ast$ a characterstic density. These values are fixed in order
to fit the current luminosities and densities of galaxies. Then at
high z, the comoving number density follows the New Luminosity Function (NLF) $$ \Phi(L,z)dL = (
1 + z)^{2\eta} \Phi(L(1 + z)^\eta ,0) dL $$ It is seen that the value
$\eta = 1.5$ gives a fair fit to the data on high redshift galaxies. The functional form has the
following properties:
\begin{enumerate}
\item [(i)] Self--similarity as suggested by the classical Press- Schecter (1974)
prescription. 
\item [(ii)] conservation of the total comoving mass density.
\item [(iii)] evolution of the comoving number density as $(1 + z)^\eta$ and
of the knee of the function as $L_\ast(z) = L_\ast(0)(1 + z)^{-\eta}$
\end{enumerate}
   \end {section}

\begin {section} {Lensing Cross-Section For SIS Galaxies}
   The probability of a beam encountering a lensing object is governed by
the parameter F (TOG 1984)  which represents the effectiveness of
distributed cosmic matter in producing double images. The parameter F is
given by
\begin{equation}F = {16\pi^{3}\over{c
H_{0}^{3}}}n_{0}v^{4}\end{equation}

\noindent
where $ n_{0}$ is the present comoving
number density of lensing galaxies and $v$ is the line of sight
velocity dispersion of  matter in the galaxy. The Hubble constant
$H_{0}$ is normalized by $ H_{0}$ = 100 h km s$^{-1}Mpc^{-1}$,  where $0.4 < h < 1.0$. 
In this
section we evaluate F from the statistics of local galaxies, assuming
that their properties persist out to distant galaxies. The
relationship between $v$ and the luminosity is known empirically,
 i.e., $L\propto v^{4}$ for elliptical galaxies and $L\propto
v^{2.6}$ for spiral galaxies in the B band. Using the luminosity
function given by Volmerange and
Guiderdoni (1990)

\begin{equation}\Phi(L,z)dL = (1 +
z)^{2\eta}\phi_{\ast}\left[{L\over L_{\ast}}(1 +
z)^{\eta}\right]^{\alpha} exp\left[ {-L\over{L_{\ast}}}(1 +
z)^{\eta}\right]{dL\over{ L_{\ast}}}\end{equation}.
\begin{equation}
n_{L}(0) =
\int_{0}^{\infty}\Phi(L,z)dL\end{equation}

\begin{equation}\left({L\over
{L_{\ast}}} \right) = \left({v\over
v_{\ast}}\right)^{\gamma}\end{equation}

\vskip .2cm
\noindent
Using
eq.(1), eq.(2), eq.(3), eq.(4) we get 

\begin{equation} F = {16\pi^{3}\over{c
H_{0}^{3}}}\phi_\ast v_\ast^{4}\Gamma\left(\alpha + {4\over\gamma} +1\right)
\left(1 + z \right)^{\eta( 1 - {4\over\gamma})}\end{equation}

\noindent
which may be written as

\begin{equation} F = F^*(1 + z)^{\eta( 1 - {4\over\gamma})}\end{equation}

where

\begin{equation} F^* = {16\pi^{3}\over{c
H_{0}^{3}}}\phi_\ast v_\ast^{4}\Gamma\left(\alpha + {4\over\gamma} +1\right)
\end{equation}.

\noindent
here $v_\ast$ is the velocity dispersion corresponding to the
characteristic luminosity $L_\ast$ and $\gamma= 4$ for elliptical
galaxies and  $\gamma= 2.6$ 
for spiral galaxies. According to Fukugita and Turner (1991),

\begin{equation} \phi_\ast = ( 1.56 \pm 0.4)\times 10^{-2} h^{3} Mpc^{-3}\end{equation}
$$\alpha = -1.1 \pm 0.1$$

\noindent 
The values of the parameter $F^\ast$ adopted by them are summarized in Table 1.

%\begin{table}[h]
%\caption{ SIS velocity dispersion and $F^\ast$ values (using
%parameters of Fukugita and Turner (1991)).}
%\begin{center}
%\begin{tabular}{|c|c|c|c|} \hline
%Type     &  Composition     & $v_\ast (km/sec)$        & $F^\ast$ \\
%\hline\hline
%E & $ 12\pm 2\%$ & $(225)^{+12}_{-25} \times \sqrt{3/2}$ & $0.019 \pm 0.008$ \\ 
%&&& \\
%SO & $ 19\pm 4\%$ & $(206)^{+12}_{-20}\times \sqrt{3/2}$ & $0.021 \pm 0.009$ \\ 
%&&& \\
%S & $ 69\pm 4 \%$ & $144^{+8}_{-13} \pm 10$ & $0.007\pm 0.003$ \\ 
%&&& \\
%ALL & $100\%$ & & $0.047\pm 0.019$\\ \hline
%\end{tabular}
%\end{center}
%\end{table}

It has been argued that the existence of a $\sqrt{3/2}$ correction factor for the dark matter
velocities  disperion used by Fukugita and Turner is incorrect (Fukugita \& Turner 1991). Better
dynamical models (Kochanek 1993; Kochanek 1994; Brimer \& Sanders
1993) show that the assumption of  $\sqrt{3/2}$ is incorrect, and is
supported neither by galactic dynamics nor by the observed separations
of gravitational lensing. The best fit estimate  
is $v_\ast = 220 \pm 20$ km/sec  for E + SO galaxies, and it is well constrained
because the average image separation is a strong function of velocity dispersion 
 (Kochanek 1996). The new values of $F^\ast$ obtained by using  the  parameters given by Kochanek (Kochanek
1996)
are summarized in Table 2.$$\phi_\ast = (1.40 \pm 0.17)h^{3}\times 10^{-2}
 Mpc^{-3}$$.
$$\alpha = -1.00 \pm 0.15$$

\noindent

%\begin{table}[h]
%\caption{ SIS velocity dispersion and $F^\ast$ values (using
%parameters of Kochanek (1996)).}
%\begin{center}
%\begin{tabular}{|c|c|c|c|} \hline
%Type     &  Composition     & $v_\ast (km/sec)$        & $F^\ast$    \\ \hline\hline
%
%E+ SO  & $ 43\%$ & $220 \pm 20$ & $0.023 $ \\ 
%&&& \\
%S & $ 57 \%$ & $144 \pm 10$ & $0.005$ \\ 
%&&& \\
%ALL & $100\%$ & & $0.028$\\ \hline
%\end{tabular}
%\end{center}
%\end{table}

\noindent
We note that the total value of  $F^\ast$ determined here $(F^\ast =
0.028)$ is four times smaller then the value used by TOG. Nakamura and
Suto (1996) have  given the new values of $(\gamma,v_\ast)$ =
$(3.3,175\, kms^{-1})$ for E and SO galaxies and $(\gamma,v_\ast)$ = $
(2.9,126\, kms^{-1})$ for S galaxies. The morphological composition is
$\phi_{\ast E} + \phi_{\ast SO}=  0.44\phi_{\ast}$ and  $\phi_{\ast S}
= 0.56 \phi_{\ast}$ 
 where  $\phi_{\ast} = 0.26 h^{3} Mpc^{-3}$ and $\alpha = - 1.09$. The
 values of $F^\ast$ wth Nakamura and Suto parameters are summarized in Table 3. 

%\vskip.2 cm
%\noindent

%\begin{table}[h]
%\caption{ SIS velocity dispersion and $F^\ast$ values (using
%parameters of Nakamura and Suto (1996)).}  
%\begin{center}
%\begin{tabular}{|c|c|c|c|} \hline
%Type     &  Composition     & $v_\ast (km/sec)$        & $F^\ast$    \\ \hline\hline

%E+ SO  & $ 44\%$ & $175$ & $0.016 $ \\ 
%&&& \\
%S & $ 56 \%$ & $126$ & $0.005$ \\ 
%&&& \\
%ALL & $100\%$ & & $0.021$\\ \hline
%\end{tabular}
%\end{center}
%\end{table}   

   \end {section}

\begin{section} {Basic Equations For Statistical Lensing}
   To discuss the statistical properties of gravitational lenses we assume that the 
universe is well approximated by the Friedmann - Lemaitre - Robertson - Walker
(FLRW) geometry on large scale. We write the basic equations for the statistical 
lensing for two models of the lensing object; i.e. , (1) Point Masses, which are 
appropriate models for stellar mini - lensing or concentrated sources such as black holes, and (2) Singular Isothermal Sphere (SIS), which would model the matter distribution of an isolated galaxy. We use the  following notation for distances: $$ D_{OL} = d(0,z_{L}), D_{LS} = d(z_{L},z_{s}), D_{OS} = d(0,z_{S}),$$ where $d(z_{1},z_{2})$ is the angular diameter 
distance between the redshift $z_{1}$ and  $z_{2}$ , and the arguments  $z_{L}$ and $z_{S}$ are the redshifts of the lens and source respectively. The formulation and 
notation of TOG is basically followed in this paper. For $\Omega_\circ + \lambda_\circ = 1$,

\begin{equation}d(z_{1},z_{2})= {R_\circ \over (1 + z_{2})}\int_{z_{1}}
^{z_{2}}{dz\over\sqrt{\Omega_\circ(1 +z)^{3} + \lambda_\circ}},\end{equation}

\noindent
where $\lambda_\circ = \Lambda/3H^{2}_\circ$ and $\Lambda = 8 \pi G \rho_{vac}$ and
$R_{\circ}$ is the present scale factor.
\vskip .3cm 
\begin{center}
{\bf 4.1.Point Masses}
\end{center}

This model can be considered to be a good approximation for many celestial bodies like ''Jupiter'', stars, black holes and even galaxies, when the light rays from background sources pass outside the deflectors. We first define the length $a_{cr}$ which characterizes the  critical radial size of the lens such that 

\begin{equation} a_{cr}^{2} = 
{4GM \over c^{2}}{D_{OL}D_{LS} \over D_{OS}},\end{equation}
\vskip .2cm
\noindent
where M is mass of the lensing object(TOG). Then the cross - section $\sigma$ for strong lensing events as defined by TOG is given by \begin{equation}\sigma  = \pi a_{cr}^{2},\end{equation}
The differential probabilty $d\tau$ of a beam encountering a lens in traversing the path of $dz_{L}$ is given by

\begin{equation} d\tau = n_{L}(z)\sigma{cdt\over dz_{L}}
dz_{L},\end{equation}.
%\vskip .1cm
\begin{equation} d\tau = {3 \over 2}\Omega_{L}(0)(1 + z_{L})^3\left({D_{OL}D_{LS} \over R_\circ D_{OS}}\right){1 \over R_\circ}{cdt \over dz_{L}}dz_{L},\end{equation} 
\vskip .3cm
\noindent
where $n_{L}(z) = n_{L}(0)(1 + z_{L})^{3}$ and $n_{L}(0)$ is the comoving number density and $\Omega_{L}(0)= 8\pi G M n_{L}(0)/3H^{2}_{\circ}$ is the lens density parameter
which is the ratio of the local lens density to the critical density. The quantity 
$cdt/dz_{L}$ is calculated in the FLRW geometry to be 

\begin{equation} {cdt \over dz_{L}} = {R_\circ \over (1 + z_{L})}{1 \over \sqrt{\Omega_\circ(1 + z_{L})^{3} + (1 -
\Omega_\circ -\lambda_\circ)(1 + z_{L})^{2} + \lambda_\circ}},\end{equation} 

\vskip .2cm
\noindent
$R_\circ$ is the Hubble distance $(R_\circ = c/H_\circ) $ ,  $ t$  stands for the look back time and  $\Omega_\circ$ is the total mass density of the universe. By integrating the differential probabilty along the line of sight to the source , we obtain the total probabilty

\begin{equation} \tau(z_{s})= \int_{0}^{z_{s}}{d\tau \over dz_{L}}dz_{L} ,\end{equation}
\vskip .3in
\begin{center}
{\bf 4.2.Singular Isothermal Spheres }
\end{center}

The singular isothermal sphere (SIS) provides us with a reasonable approximation to 
account for the lensing properties of a real galaxy. The lens model is characterized
by the one dimensional velosity dispersion $v$. The deflection angle for all impact parameters is given by $\alpha = 4\pi v^{2}/c^{2}$. The lens produces two images if the angular position of the source is less than the critical angle $\beta_{cr}$, analogous to the critical radius in the previous subsection, which is the deflection of a beam passing  at any radius through a SIS:

\begin{equation} \beta_{cr} = \alpha D_{LS}/D_{OS} ,\end{equation} 

\noindent
Then the critical impact parameter is defined by $a_{cr} = 
D_{OL}\beta_{cr}$ and the cross- section is given by 

\begin{equation}\sigma = \pi a_{cr}^{2} = 16{\pi}^{3}\left({v \over c}\right)^{4}\left({D_{OL}D_{LS}\over D_{OS}}\right)^{2} ,\end{equation}

\noindent
The differential probability  of a lensing event is given by using eqs.(2), (5), (12), (17),$$d\tau = n_{L}(0) (1 + z_{L})^{3}\sigma {cdt \over dz_{L}} dz_{L}$$.$$d\tau = F(1 + z_{L})^{3}\left({D_{OL}D_{LS}\over R_\circ D_{OS}}\right)^{2}
{1\over R_\circ}{cdt \over dz_{L}} dz_{L}$$.
\begin{eqnarray}
d\tau &=& {16\pi^{3}\over{c
H_{0}^{3}}}\,\phi_\ast\, v_\ast^{4}\Gamma\left(\alpha + {4\over\gamma} +1\right)(1 + z)^{\eta\,( 1 - {4\over\gamma})} \nonumber\\
& & \nonumber\\
&&\times\,(1 + z_{L})^{3}
\left({D_{OL}D_{LS}\over R_\circ D_{OS}}\right)^{2}{1\over R_\circ}{cdt \over dz_{L}} dz_{L}\end{eqnarray} 

\noindent
It turns out that for $\gamma = 4$ the differential probabilty is the same for both PSLF and NLF.
The total probabilty  obtained by integrating the differential probability 
along the the line of sight to the source as in eq.(15), is plotted in Fig.1. 
with both PSLF and NLF for $\Omega_\circ$ = 0.1 and  $\lambda_\circ$ = 0.9 .

   \end {section}

\begin{section} {The Quasar Luminosity Function }
   Boyle, Shanks \& Peterson (1988) (BSP) have published a sample of 420 faint ultraviolet - 
excess quasars that extends from z = 0.3 to z = 2.2 and is complete for magnitude brighter than 20.9 mag. Using this sample, they proposed a quasar luminosity function of the form

\begin{equation} \Phi(M_{B},z) =  {\Phi^{\ast}\over {10^{0.4[M_{B}-M_{B}(z)](\alpha +1)} +10^{0.4[M_{B}-M_{B}(z)](\beta +1)}}}{M\-pc}^{-3}mag^{-1}\end{equation}

\noindent
where $M_{B}(z)$,  the magnitude at which there is a turnover
in the power law slope, varies as  a function of quasar redshift :

\begin{equation}
M_{B}(z) = M_{B}^{\ast} - 2.5\, k_{L}\, Log(1 + z),\end{equation} 

\vskip .3cm
\noindent
Here
$M_{B}^{\ast}$ and $k_{L}$, as well as $\alpha$ and $\beta$ from
eq.(19) are constant given by the best fit of the parameters to the
data and $\Phi^{\ast}$ is the normalizing factor. We have used model B
from BSP, a pure power law evolution that fits the data very well. For
this model, the values of the constants are $M_{B}^{\ast} = -22.42,\,
 k_{L} =3.15,\, \alpha = -3.79,\, \beta = -1.44$ .   

%\vskip .3cm
%\noindent
%{\underbar{ How to fix the value of $\Phi^*$}} : Integrate the Quasar Luminosity 
%Function over redshift and apparent magnitude and fix the value of $\Phi^*$ in such%a way that it gives back the total number of orignal quasars in the sample. 

   \end {section}

\begin{section} {The Number Of Lensed Quasars }
   Besides depending upon the number density of galaxies and their
properties, the number of quasars gravitationally lensed also depends
upon the number of unlensed quasars. 
This is because of the amplification bias i.e., the brightening of
lensed quasars, the number of quasars a few  magnitudes fainter than
the detection limit is important in predicting the number of lensed
quasars in a flux limited sample. 
Fukugita \& Turner (1991) calculated the number of lensed quasars
expected for $\lambda_{\circ}$ cosmology taking into the account the
amplification bias. 
The number--magnitude counts of the lensed quasars are given by the
relation 

\begin{equation} N_{LQ}(m) = \tau\int_{0}^{\infty}N_{Q}( m +
\Delta)P(\Delta)d\Delta,\end{equation}

\vskip .2cm
\noindent
where $ P(\Delta)d\Delta$ is the
probabilty that lensing will cause the magnitude of the quasar
(images) to decrease by $\Delta$; i,e., the images are brighter by a
factor $ A$; $\Delta = 2.5\, Log A$ .

\begin{equation}  N_{LQ} =
\int_{0}^{\infty}P(\Delta)d\Delta \int_{z1}^{z2} dz\, \tau(z)\,{dV\over dz}
\int_{M1}^{M2} \Phi(M_{B},z)dM_{B},\end{equation} 

\vskip .2cm
\noindent
where $P(\Delta)\,d\Delta = 7.37\, 10^{-0.8\Delta} d\Delta$ and  $dV/dz$ is
the comoving volume which is given by

\begin{equation}  {dV\over dz} = {4 \pi\, d_{L}^2\,
c \over{H_{\circ}(1 +z)^{2}\sqrt{\Omega_{\circ}(1 + z)^{3} +
\lambda_{\circ}- (\Omega_{\circ} + \lambda_{\circ} - 1)(1 + z)^2}}},\end{equation}

\vskip .2 cm
\noindent
where $d_{L}$ is the luminosity distance. The $N_{LQ}$ calculated in BSP sample by using PSLF and NLF for galaxy distribution are summarized in Table 4.
\noindent
The result is plotted in Fig. 2
%\vskip .3cm
%\noindent
%Table 4. Number of lensed quasars $N_{LQ}$ in BSP sample.

%\begin{table}[h]
%\begin{center}
%\begin{tabular}{|c|c|c|c|} \hline
%$\Omega_\circ + \lambda_\circ $   & $N_{LQ}$ with PSLF& $N_{LQ}$ with NLF& $N_{LQ}$% with PSLF    \\ 
% & $ F^{\ast}$ = 0.028 & $ F^{\ast}$ = 0.021 &  $ F^{\ast}$ = 0.021 \\ \hline\hlin%e

% $1.0 + 0.0 $ & $0.58$ & $0.38 $ & $0.43$ \\
%
% $0.9 + 0.1 $ & $0.65$ & $0.43 $ & $0.49$ \\
%
% $0.8 + 0.2 $ & $0.72$ & $0.47 $ & $0.54$ \\
%
% $0.7 + 0.3 $ & $0.81$ & $0.53 $ & $0.61$ \\

% $0.6 + 0.4 $ & $0.92$ & $0.60 $ & $0.69$ \\

% $0.5 + 0.5 $ & $1.06$ & $0.69 $ & $0.79$ \\

% $0.4 + 0.6 $ & $1.29$ & $0.83 $ & $0.97$ \\

% $0.3 + 0.7 $ & $1.59$ & $1.02 $ & $1.19$ \\

% $0.2 + 0.8 $ & $2.11$ & $1.35 $ & $1.58$ \\

% $0.1 + 0.9 $ & $3.13$ & $1.98 $ & $2.35$ \\

% $0.05 + 0.95 $ & $4.13$ & $2.59 $ & $3.09$ \\

% $0.01 + 0.99 $ & $5.83$ & $3.61 $ & $4.37$ \\

% $0.0 + 1.0 $ & $6.51$ & $4.02 $ & $4.88$ \\ \hline

%\end{tabular}
%\end{center}
%\end{table}
%\noindent
%The result is plotted in Fig. 2.

   \end {section}

\begin{section} {Conclusion}
   An upper bound to $\lambda$, results from the comparison of the statistics of quasars observed to be gravitationally lensed by intervening galaxies, with the predictions of flat cosmological models with a nonzero cosmological constant.  A flat cosmology with cosmological constant tends to produce more gravitationally lensed quasars then does such a cosmology with zero  $\lambda$ because a large  $\lambda$ increases the volume per unit redshift of the universe at high redshift. This means that the relative number of lensed quasars for a fixed comoving number density of galaxies increases rapidly with increasing  $\lambda$. Turning this around it is possible to use the observed probabilty of lensing to constrain  $\lambda$. This method has been used by various authors (Turner 1990, Fukugita $\&$ Turner 1991, Fukugita et al. 1992) and recently Kochanek has put limit on  $\lambda$ i.e.,  $\lambda < 0.65$ (Kochanek 1996).

 However, limited observational data and the possibility that evolution effects on the population of lensing galaxies have not been properly taken into account suggest that this limit is still uncertain. We have tried to put limit on  $\lambda$ by taking evolutionary effect of galaxies. The important result can be described as follows: The probabilty of a beam encountering with lens event has been reduced in an evolutionary picture (NLF) and as a consequence of it, the evolutionary model permits higher value of $\lambda$. This can be explained as follows, since there is no observed lensed quasars in BSP sample. Taking a typical uncertainity of $\approx \pm 1$ in the number of lensed quasars, PSLF put limit on $\lambda$ i.e., $\lambda < 0.6$ with $F^* = 0.021$ whereas  NLF gives  $\lambda < 0.7$ with $F^* = 0.021$. And NLF also gives  $\lambda < 0.47$ with $F^* = 0.028$.

   \end {section}

\begin{section}*{Acknowledgements}
   
This work is funded in part by DST, Government of India, under project (SP/ S2/ K- 06/91). One of the authors (Deepak Jain) is thankful to GATE for  fellowship and also
thankful to Harvinder K. Jassal for useful discussions.
   \end {section}
   \vfill
   \eject

\vfill
\eject
%\begin{center}
%\input {fig2.tex}
%\end{center}

\begin{section}*{Figure Captions}
   Fig. 1 \, $z_s$ \,  Vs \,  $\tau / F^* $ 

\noindent
Fig. 2  \,$\lambda$  \, Vs  \, $N_{LQ}$
   \end {section}
   \vfill
   \eject

\begin{section}*{Table 1.  SIS velocity dispersion and $F^\ast$ values (using
parameters of Fukugita and Turner (1991)).}
   
\begin{table}[h]
%\caption{ SIS velocity dispersion and $F^\ast$ values (using
%parameters of Fukugita and Turner (1991)).}
\begin{center}
\begin{tabular}{|c|c|c|c|} \hline
Type     &  Composition     & $v_\ast (km/sec)$        & $F^\ast$ \\
\hline\hline
E & $ 12\pm 2\%$ & $(225)^{+12}_{-25} \times \sqrt{3/2}$ & $0.019 \pm 0.008$ \\ 
&&& \\
SO & $ 19\pm 4\%$ & $(206)^{+12}_{-20}\times \sqrt{3/2}$ & $0.021 \pm 0.009$ \\ 
&&& \\
S & $ 69\pm 4 \%$ & $144^{+8}_{-13} \pm 10$ & $0.007\pm 0.003$ \\ 
&&& \\
ALL & $100\%$ & & $0.047\pm 0.019$\\ \hline
\end{tabular}
\end{center}
\end{table}

   \end {section}
   \vfill
   \eject
\begin{section}*{Table 2.  SIS velocity dispersion and $F^\ast$ values (using
parameters of Kochanek (1996)).}
   \begin{table}[h]
%\caption{ SIS velocity dispersion and $F^\ast$ values (using
%parameters of Kochanek (1996)).}
\begin{center}
\begin{tabular}{|c|c|c|c|} \hline
Type     &  Composition     & $v_\ast (km/sec)$        & $F^\ast$    \\ \hline\hline

E+ SO  & $ 43\%$ & $220 \pm 20$ & $0.023 $ \\ 
&&& \\
S & $ 57 \%$ & $144 \pm 10$ & $0.005$ \\ 
&&& \\
ALL & $100\%$ & & $0.028$\\ \hline
\end{tabular}
\end{center}
\end{table}
   \end {section}
   \vfill
   \eject
\begin{section}*{Table 3. SIS velocity dispersion and $F^\ast$ values (using
parameters of Nakamura and Suto (1996)).}
   \begin{table}[h]
%\caption{ SIS velocity dispersion and $F^\ast$ values (using
%parameters of Nakamura and Suto (1996)).}  
\begin{center}
\begin{tabular}{|c|c|c|c|} \hline
Type     &  Composition     & $v_\ast (km/sec)$        & $F^\ast$    \\ \hline\hline

E+ SO  & $ 44\%$ & $175$ & $0.016 $ \\ 
&&& \\
S & $ 56 \%$ & $126$ & $0.005$ \\ 
&&& \\
ALL & $100\%$ & & $0.021$\\ \hline
\end{tabular}
\end{center}
\end{table}
   \end {section}
   \vfill
   \eject

\begin{section}*{Table 4. Number of lensed quasars $N_{LQ}$ in BSP sample.}
   \begin{table}[h]
\begin{center}
\begin{tabular}{|c|c|c|c|} \hline
$\Omega_\circ + \lambda_\circ $   & $N_{LQ}$ with NLF& $N_{LQ}$ with NLF& $N_{LQ}$ with PSLF    \\ 
 & $ F^{\ast}$ = 0.028 & $ F^{\ast}$ = 0.021 &  $ F^{\ast}$ = 0.021 \\ \hline\hline

 $1.0 + 0.0 $ & $0.58$ & $0.38 $ & $0.43$ \\

 $0.9 + 0.1 $ & $0.65$ & $0.43 $ & $0.49$ \\

 $0.8 + 0.2 $ & $0.72$ & $0.47 $ & $0.54$ \\

 $0.7 + 0.3 $ & $0.81$ & $0.53 $ & $0.61$ \\

 $0.6 + 0.4 $ & $0.92$ & $0.60 $ & $0.69$ \\

 $0.5 + 0.5 $ & $1.06$ & $0.69 $ & $0.79$ \\

 $0.4 + 0.6 $ & $1.29$ & $0.83 $ & $0.97$ \\

 $0.3 + 0.7 $ & $1.59$ & $1.02 $ & $1.19$ \\

 $0.2 + 0.8 $ & $2.11$ & $1.35 $ & $1.58$ \\

 $0.1 + 0.9 $ & $3.13$ & $1.98 $ & $2.35$ \\

 $0.05 + 0.95 $ & $4.13$ & $2.59 $ & $3.09$ \\

 $0.01 + 0.99 $ & $5.83$ & $3.61 $ & $4.37$ \\

 $0.0 + 1.0 $ & $6.51$ & $4.02 $ & $4.88$ \\ \hline

\end{tabular}
\end{center}
\end{table}
   \end {section}
   \vfill
   \eject
%\documentstyle [12pt]{article}
%\begin{document}
\bf
\large
\pagestyle{empty}
% GNUPLOT: LaTeX picture
\setlength{\unitlength}{0.240900pt}
\ifx\plotpoint\undefined\newsavebox{\plotpoint}\fi
\sbox{\plotpoint}{\rule[-0.200pt]{0.400pt}{0.400pt}}%
\begin{picture}(2250,1575)(400,0)
\font\gnuplot=cmr10 at 10pt
\gnuplot
\sbox{\plotpoint}{\rule[-0.200pt]{0.400pt}{0.400pt}}%
\put(220.0,113.0){\rule[-0.200pt]{473.609pt}{0.400pt}}
\put(220.0,113.0){\rule[-0.200pt]{0.400pt}{335.815pt}}
\put(220.0,113.0){\rule[-0.200pt]{4.818pt}{0.400pt}}
\put(198,113){\makebox(0,0)[r]{0}}
\put(2166.0,113.0){\rule[-0.200pt]{4.818pt}{0.400pt}}
\put(220.0,345.0){\rule[-0.200pt]{4.818pt}{0.400pt}}
\put(198,345){\makebox(0,0)[r]{0.05}}
\put(2166.0,345.0){\rule[-0.200pt]{4.818pt}{0.400pt}}
\put(220.0,578.0){\rule[-0.200pt]{4.818pt}{0.400pt}}
\put(198,578){\makebox(0,0)[r]{0.1}}
\put(2166.0,578.0){\rule[-0.200pt]{4.818pt}{0.400pt}}
\put(220.0,810.0){\rule[-0.200pt]{4.818pt}{0.400pt}}
\put(198,810){\makebox(0,0)[r]{0.15}}
\put(2166.0,810.0){\rule[-0.200pt]{4.818pt}{0.400pt}}
\put(220.0,1042.0){\rule[-0.200pt]{4.818pt}{0.400pt}}
\put(198,1042){\makebox(0,0)[r]{0.2}}
\put(2166.0,1042.0){\rule[-0.200pt]{4.818pt}{0.400pt}}
\put(220.0,1275.0){\rule[-0.200pt]{4.818pt}{0.400pt}}
\put(198,1275){\makebox(0,0)[r]{0.25}}
\put(2166.0,1275.0){\rule[-0.200pt]{4.818pt}{0.400pt}}
\put(220.0,1507.0){\rule[-0.200pt]{4.818pt}{0.400pt}}
\put(198,1507){\makebox(0,0)[r]{0.3}}
\put(2166.0,1507.0){\rule[-0.200pt]{4.818pt}{0.400pt}}
\put(220.0,113.0){\rule[-0.200pt]{0.400pt}{4.818pt}}
\put(220,68){\makebox(0,0){0}}
\put(220.0,1487.0){\rule[-0.200pt]{0.400pt}{4.818pt}}
\put(548.0,113.0){\rule[-0.200pt]{0.400pt}{4.818pt}}
\put(548,68){\makebox(0,0){0.5}}
\put(548.0,1487.0){\rule[-0.200pt]{0.400pt}{4.818pt}}
\put(875.0,113.0){\rule[-0.200pt]{0.400pt}{4.818pt}}
\put(875,68){\makebox(0,0){1}}
\put(875.0,1487.0){\rule[-0.200pt]{0.400pt}{4.818pt}}
\put(1203.0,113.0){\rule[-0.200pt]{0.400pt}{4.818pt}}
\put(1203,68){\makebox(0,0){1.5}}
\put(1203.0,1487.0){\rule[-0.200pt]{0.400pt}{4.818pt}}
\put(1531.0,113.0){\rule[-0.200pt]{0.400pt}{4.818pt}}
\put(1531,68){\makebox(0,0){2}}
\put(1531.0,1487.0){\rule[-0.200pt]{0.400pt}{4.818pt}}
\put(1858.0,113.0){\rule[-0.200pt]{0.400pt}{4.818pt}}
\put(1858,68){\makebox(0,0){2.5}}
\put(1858.0,1487.0){\rule[-0.200pt]{0.400pt}{4.818pt}}
\put(2186.0,113.0){\rule[-0.200pt]{0.400pt}{4.818pt}}
\put(2186,68){\makebox(0,0){3}}
\put(2186.0,1487.0){\rule[-0.200pt]{0.400pt}{4.818pt}}
\put(220.0,113.0){\rule[-0.200pt]{473.609pt}{0.400pt}}
\put(2186.0,113.0){\rule[-0.200pt]{0.400pt}{335.815pt}}
\put(220.0,1507.0){\rule[-0.200pt]{473.609pt}{0.400pt}}
\put(45,810){\makebox(0,0){$\tau/F^{*}$}}
\put(1203,23){\makebox(0,0){$z_{s}$}}
\put(1203,1552){\makebox(0,0){Fig. 1}}
\put(1334,949){\makebox(0,0)[l]{With PSLF}}
\put(1695,461){\makebox(0,0)[l]{With NLF}}
\put(220.0,113.0){\rule[-0.200pt]{0.400pt}{335.815pt}}
\put(1531,949){\vector(1,0){281}}
\put(1695,461){\vector(-1,0){328}}
\put(286,113){\usebox{\plotpoint}}
\put(286,112.67){\rule{15.658pt}{0.400pt}}
\multiput(286.00,112.17)(32.500,1.000){2}{\rule{7.829pt}{0.400pt}}
\multiput(351.00,114.61)(14.528,0.447){3}{\rule{8.900pt}{0.108pt}}
\multiput(351.00,113.17)(47.528,3.000){2}{\rule{4.450pt}{0.400pt}}
\multiput(417.00,117.59)(7.167,0.477){7}{\rule{5.300pt}{0.115pt}}
\multiput(417.00,116.17)(54.000,5.000){2}{\rule{2.650pt}{0.400pt}}
\multiput(482.00,122.59)(4.324,0.488){13}{\rule{3.400pt}{0.117pt}}
\multiput(482.00,121.17)(58.943,8.000){2}{\rule{1.700pt}{0.400pt}}
\multiput(548.00,130.58)(3.043,0.492){19}{\rule{2.464pt}{0.118pt}}
\multiput(548.00,129.17)(59.887,11.000){2}{\rule{1.232pt}{0.400pt}}
\multiput(613.00,141.58)(2.241,0.494){27}{\rule{1.860pt}{0.119pt}}
\multiput(613.00,140.17)(62.139,15.000){2}{\rule{0.930pt}{0.400pt}}
\multiput(679.00,156.58)(1.831,0.495){33}{\rule{1.544pt}{0.119pt}}
\multiput(679.00,155.17)(61.794,18.000){2}{\rule{0.772pt}{0.400pt}}
\multiput(744.00,174.58)(1.448,0.496){43}{\rule{1.248pt}{0.120pt}}
\multiput(744.00,173.17)(63.410,23.000){2}{\rule{0.624pt}{0.400pt}}
\multiput(810.00,197.58)(1.212,0.497){51}{\rule{1.063pt}{0.120pt}}
\multiput(810.00,196.17)(62.794,27.000){2}{\rule{0.531pt}{0.400pt}}
\multiput(875.00,224.58)(1.106,0.497){57}{\rule{0.980pt}{0.120pt}}
\multiput(875.00,223.17)(63.966,30.000){2}{\rule{0.490pt}{0.400pt}}
\multiput(941.00,254.58)(0.960,0.498){65}{\rule{0.865pt}{0.120pt}}
\multiput(941.00,253.17)(63.205,34.000){2}{\rule{0.432pt}{0.400pt}}
\multiput(1006.00,288.58)(0.871,0.498){73}{\rule{0.795pt}{0.120pt}}
\multiput(1006.00,287.17)(64.350,38.000){2}{\rule{0.397pt}{0.400pt}}
\multiput(1072.00,326.58)(0.794,0.498){79}{\rule{0.734pt}{0.120pt}}
\multiput(1072.00,325.17)(63.476,41.000){2}{\rule{0.367pt}{0.400pt}}
\multiput(1137.00,367.58)(0.751,0.498){85}{\rule{0.700pt}{0.120pt}}
\multiput(1137.00,366.17)(64.547,44.000){2}{\rule{0.350pt}{0.400pt}}
\multiput(1203.00,411.58)(0.688,0.498){93}{\rule{0.650pt}{0.120pt}}
\multiput(1203.00,410.17)(64.651,48.000){2}{\rule{0.325pt}{0.400pt}}
\multiput(1269.00,459.58)(0.650,0.498){97}{\rule{0.620pt}{0.120pt}}
\multiput(1269.00,458.17)(63.713,50.000){2}{\rule{0.310pt}{0.400pt}}
\multiput(1334.00,509.58)(0.635,0.498){101}{\rule{0.608pt}{0.120pt}}
\multiput(1334.00,508.17)(64.739,52.000){2}{\rule{0.304pt}{0.400pt}}
\multiput(1400.00,561.58)(0.591,0.499){107}{\rule{0.573pt}{0.120pt}}
\multiput(1400.00,560.17)(63.811,55.000){2}{\rule{0.286pt}{0.400pt}}
\multiput(1465.00,616.58)(0.589,0.499){109}{\rule{0.571pt}{0.120pt}}
\multiput(1465.00,615.17)(64.814,56.000){2}{\rule{0.286pt}{0.400pt}}
\multiput(1531.00,672.58)(0.550,0.499){115}{\rule{0.541pt}{0.120pt}}
\multiput(1531.00,671.17)(63.878,59.000){2}{\rule{0.270pt}{0.400pt}}
\multiput(1596.00,731.58)(0.550,0.499){117}{\rule{0.540pt}{0.120pt}}
\multiput(1596.00,730.17)(64.879,60.000){2}{\rule{0.270pt}{0.400pt}}
\multiput(1662.00,791.58)(0.524,0.499){121}{\rule{0.519pt}{0.120pt}}
\multiput(1662.00,790.17)(63.922,62.000){2}{\rule{0.260pt}{0.400pt}}
\multiput(1727.00,853.58)(0.532,0.499){121}{\rule{0.526pt}{0.120pt}}
\multiput(1727.00,852.17)(64.909,62.000){2}{\rule{0.263pt}{0.400pt}}
\multiput(1793.00,915.58)(0.507,0.499){125}{\rule{0.506pt}{0.120pt}}
\multiput(1793.00,914.17)(63.949,64.000){2}{\rule{0.253pt}{0.400pt}}
\multiput(1858.00,979.58)(0.507,0.499){127}{\rule{0.506pt}{0.120pt}}
\multiput(1858.00,978.17)(64.949,65.000){2}{\rule{0.253pt}{0.400pt}}
\multiput(1924.00,1044.58)(0.499,0.499){127}{\rule{0.500pt}{0.120pt}}
\multiput(1924.00,1043.17)(63.962,65.000){2}{\rule{0.250pt}{0.400pt}}
\multiput(1989.00,1109.58)(0.499,0.499){129}{\rule{0.500pt}{0.120pt}}
\multiput(1989.00,1108.17)(64.962,66.000){2}{\rule{0.250pt}{0.400pt}}
\multiput(2055.58,1175.00)(0.499,0.515){127}{\rule{0.120pt}{0.512pt}}
\multiput(2054.17,1175.00)(65.000,65.937){2}{\rule{0.400pt}{0.256pt}}
\multiput(2120.58,1242.00)(0.499,0.507){129}{\rule{0.120pt}{0.506pt}}
\multiput(2119.17,1242.00)(66.000,65.950){2}{\rule{0.400pt}{0.253pt}}
\put(286,113){\usebox{\plotpoint}}
\multiput(286,113)(20.753,0.319){4}{\usebox{\plotpoint}}
\multiput(351,114)(20.734,0.942){3}{\usebox{\plotpoint}}
\multiput(417,117)(20.716,1.275){3}{\usebox{\plotpoint}}
\multiput(482,121)(20.605,2.498){3}{\usebox{\plotpoint}}
\multiput(548,129)(20.514,3.156){3}{\usebox{\plotpoint}}
\multiput(613,139)(20.364,4.011){4}{\usebox{\plotpoint}}
\multiput(679,152)(20.080,5.252){3}{\usebox{\plotpoint}}
\multiput(744,169)(19.864,6.019){3}{\usebox{\plotpoint}}
\multiput(810,189)(19.567,6.924){3}{\usebox{\plotpoint}}
\multiput(875,212)(19.311,7.607){4}{\usebox{\plotpoint}}
\multiput(941,238)(18.955,8.457){3}{\usebox{\plotpoint}}
\multiput(1006,267)(18.676,9.055){4}{\usebox{\plotpoint}}
\multiput(1072,299)(18.391,9.620){3}{\usebox{\plotpoint}}
\multiput(1137,333)(18.105,10.150){4}{\usebox{\plotpoint}}
\multiput(1203,370)(17.869,10.559){4}{\usebox{\plotpoint}}
\multiput(1269,409)(17.677,10.878){3}{\usebox{\plotpoint}}
\multiput(1334,449)(17.390,11.330){4}{\usebox{\plotpoint}}
\multiput(1400,492)(17.188,11.635){4}{\usebox{\plotpoint}}
\multiput(1465,536)(17.149,11.692){4}{\usebox{\plotpoint}}
\multiput(1531,581)(16.942,11.990){4}{\usebox{\plotpoint}}
\multiput(1596,627)(16.907,12.040){4}{\usebox{\plotpoint}}
\multiput(1662,674)(16.574,12.494){3}{\usebox{\plotpoint}}
\multiput(1727,723)(16.786,12.208){4}{\usebox{\plotpoint}}
\multiput(1793,771)(16.451,12.655){4}{\usebox{\plotpoint}}
\multiput(1858,821)(16.665,12.372){4}{\usebox{\plotpoint}}
\multiput(1924,870)(16.451,12.655){4}{\usebox{\plotpoint}}
\multiput(1989,920)(16.424,12.691){4}{\usebox{\plotpoint}}
\multiput(2055,971)(16.451,12.655){4}{\usebox{\plotpoint}}
\multiput(2120,1021)(16.424,12.691){4}{\usebox{\plotpoint}}
\put(2186,1072){\usebox{\plotpoint}}
\end{picture}
%\end{document}
%\documentstyle [12pt]{article}
%\begin{document}
\bf
\large
\pagestyle{empty}
% GNUPLOT: LaTeX picture
\setlength{\unitlength}{0.240900pt}
\ifx\plotpoint\undefined\newsavebox{\plotpoint}\fi
\sbox{\plotpoint}{\rule[-0.200pt]{0.400pt}{0.400pt}}%
\begin{picture}(2250,1575)(400,0)
\font\gnuplot=cmr10 at 10pt
\gnuplot
\sbox{\plotpoint}{\rule[-0.200pt]{0.400pt}{0.400pt}}%
\put(220.0,113.0){\rule[-0.200pt]{473.609pt}{0.400pt}}
\put(220.0,113.0){\rule[-0.200pt]{0.400pt}{335.815pt}}
\put(220.0,113.0){\rule[-0.200pt]{4.818pt}{0.400pt}}
\put(198,113){\makebox(0,0)[r]{0}}
\put(2166.0,113.0){\rule[-0.200pt]{4.818pt}{0.400pt}}
\put(220.0,312.0){\rule[-0.200pt]{4.818pt}{0.400pt}}
\put(198,312){\makebox(0,0)[r]{1}}
\put(2166.0,312.0){\rule[-0.200pt]{4.818pt}{0.400pt}}
\put(220.0,511.0){\rule[-0.200pt]{4.818pt}{0.400pt}}
\put(198,511){\makebox(0,0)[r]{2}}
\put(2166.0,511.0){\rule[-0.200pt]{4.818pt}{0.400pt}}
\put(220.0,710.0){\rule[-0.200pt]{4.818pt}{0.400pt}}
\put(198,710){\makebox(0,0)[r]{3}}
\put(2166.0,710.0){\rule[-0.200pt]{4.818pt}{0.400pt}}
\put(220.0,910.0){\rule[-0.200pt]{4.818pt}{0.400pt}}
\put(198,910){\makebox(0,0)[r]{4}}
\put(2166.0,910.0){\rule[-0.200pt]{4.818pt}{0.400pt}}
\put(220.0,1109.0){\rule[-0.200pt]{4.818pt}{0.400pt}}
\put(198,1109){\makebox(0,0)[r]{5}}
\put(2166.0,1109.0){\rule[-0.200pt]{4.818pt}{0.400pt}}
\put(220.0,1308.0){\rule[-0.200pt]{4.818pt}{0.400pt}}
\put(198,1308){\makebox(0,0)[r]{6}}
\put(2166.0,1308.0){\rule[-0.200pt]{4.818pt}{0.400pt}}
\put(220.0,1507.0){\rule[-0.200pt]{4.818pt}{0.400pt}}
\put(198,1507){\makebox(0,0)[r]{7}}
\put(2166.0,1507.0){\rule[-0.200pt]{4.818pt}{0.400pt}}
\put(220.0,113.0){\rule[-0.200pt]{0.400pt}{4.818pt}}
\put(220,68){\makebox(0,0){0}}
\put(220.0,1487.0){\rule[-0.200pt]{0.400pt}{4.818pt}}
\put(613.0,113.0){\rule[-0.200pt]{0.400pt}{4.818pt}}
\put(613,68){\makebox(0,0){0.2}}
\put(613.0,1487.0){\rule[-0.200pt]{0.400pt}{4.818pt}}
\put(1006.0,113.0){\rule[-0.200pt]{0.400pt}{4.818pt}}
\put(1006,68){\makebox(0,0){0.4}}
\put(1006.0,1487.0){\rule[-0.200pt]{0.400pt}{4.818pt}}
\put(1400.0,113.0){\rule[-0.200pt]{0.400pt}{4.818pt}}
\put(1400,68){\makebox(0,0){0.6}}
\put(1400.0,1487.0){\rule[-0.200pt]{0.400pt}{4.818pt}}
\put(1793.0,113.0){\rule[-0.200pt]{0.400pt}{4.818pt}}
\put(1793,68){\makebox(0,0){0.8}}
\put(1793.0,1487.0){\rule[-0.200pt]{0.400pt}{4.818pt}}
\put(2186.0,113.0){\rule[-0.200pt]{0.400pt}{4.818pt}}
\put(2186,68){\makebox(0,0){1}}
\put(2186.0,1487.0){\rule[-0.200pt]{0.400pt}{4.818pt}}
\put(220.0,113.0){\rule[-0.200pt]{473.609pt}{0.400pt}}
\put(2186.0,113.0){\rule[-0.200pt]{0.400pt}{335.815pt}}
\put(220.0,1507.0){\rule[-0.200pt]{473.609pt}{0.400pt}}
\put(45,810){\makebox(0,0){$N_{LQ}$}}
\put(1203,23){\makebox(0,0){$\lambda$}}
\put(1203,1552){\makebox(0,0){Fig. 2}}
\put(1793,1109){\makebox(0,0)[r]{$F^{*}=0.028$}}
\put(1793,910){\makebox(0,0)[r]{$F^{*}=0.021$}}
\put(1793,710){\makebox(0,0)[r]{$F^{*}=0.021$}}
\put(220.0,113.0){\rule[-0.200pt]{0.400pt}{335.815pt}}
\put(1793,1109){\vector(1,0){334}}
\put(1793,910){\vector(1,0){354}}
\put(1793,710){\vector(1,0){340}}
\put(2056,1442){\makebox(0,0)[r]{with NLF}}
\put(2078.0,1442.0){\rule[-0.200pt]{15.899pt}{0.400pt}}
\put(220,229){\usebox{\plotpoint}}
\multiput(220.00,229.58)(7.792,0.493){23}{\rule{6.162pt}{0.119pt}}
\multiput(220.00,228.17)(184.211,13.000){2}{\rule{3.081pt}{0.400pt}}
\multiput(417.00,242.58)(7.181,0.494){25}{\rule{5.700pt}{0.119pt}}
\multiput(417.00,241.17)(184.169,14.000){2}{\rule{2.850pt}{0.400pt}}
\multiput(613.00,256.58)(5.575,0.495){33}{\rule{4.478pt}{0.119pt}}
\multiput(613.00,255.17)(187.706,18.000){2}{\rule{2.239pt}{0.400pt}}
\multiput(810.00,274.58)(4.520,0.496){41}{\rule{3.664pt}{0.120pt}}
\multiput(810.00,273.17)(188.396,22.000){2}{\rule{1.832pt}{0.400pt}}
\multiput(1006.00,296.58)(3.556,0.497){53}{\rule{2.914pt}{0.120pt}}
\multiput(1006.00,295.17)(190.951,28.000){2}{\rule{1.457pt}{0.400pt}}
\multiput(1203.00,324.58)(2.153,0.498){89}{\rule{1.813pt}{0.120pt}}
\multiput(1203.00,323.17)(193.237,46.000){2}{\rule{0.907pt}{0.400pt}}
\multiput(1400.00,370.58)(1.639,0.499){117}{\rule{1.407pt}{0.120pt}}
\multiput(1400.00,369.17)(193.080,60.000){2}{\rule{0.703pt}{0.400pt}}
\multiput(1596.00,430.58)(1.013,0.498){95}{\rule{0.908pt}{0.120pt}}
\multiput(1596.00,429.17)(97.115,49.000){2}{\rule{0.454pt}{0.400pt}}
\multiput(1695.00,479.58)(0.909,0.498){105}{\rule{0.826pt}{0.120pt}}
\multiput(1695.00,478.17)(96.286,54.000){2}{\rule{0.413pt}{0.400pt}}
\multiput(1793.00,533.58)(0.628,0.499){153}{\rule{0.603pt}{0.120pt}}
\multiput(1793.00,532.17)(96.749,78.000){2}{\rule{0.301pt}{0.400pt}}
\multiput(1891.58,611.00)(0.498,0.512){75}{\rule{0.120pt}{0.510pt}}
\multiput(1890.17,611.00)(39.000,38.941){2}{\rule{0.400pt}{0.255pt}}
\multiput(1930.58,651.00)(0.499,0.670){115}{\rule{0.120pt}{0.636pt}}
\multiput(1929.17,651.00)(59.000,77.681){2}{\rule{0.400pt}{0.318pt}}
\multiput(1989.58,730.00)(0.499,0.866){115}{\rule{0.120pt}{0.792pt}}
\multiput(1988.17,730.00)(59.000,100.357){2}{\rule{0.400pt}{0.396pt}}
\multiput(2048.58,832.00)(0.498,1.294){77}{\rule{0.120pt}{1.130pt}}
\multiput(2047.17,832.00)(40.000,100.655){2}{\rule{0.400pt}{0.565pt}}
\multiput(2088.58,935.00)(0.495,1.974){35}{\rule{0.119pt}{1.658pt}}
\multiput(2087.17,935.00)(19.000,70.559){2}{\rule{0.400pt}{0.829pt}}
\multiput(2107.58,1009.00)(0.496,2.536){37}{\rule{0.119pt}{2.100pt}}
\multiput(2106.17,1009.00)(20.000,95.641){2}{\rule{0.400pt}{1.050pt}}
\multiput(2127.58,1109.00)(0.498,2.129){75}{\rule{0.120pt}{1.792pt}}
\multiput(2126.17,1109.00)(39.000,161.280){2}{\rule{0.400pt}{0.896pt}}
\multiput(2166.58,1274.00)(0.496,3.427){37}{\rule{0.119pt}{2.800pt}}
\multiput(2165.17,1274.00)(20.000,129.188){2}{\rule{0.400pt}{1.400pt}}
\put(2056,1397){\makebox(0,0)[r]{ with PSLF}}
\multiput(2078,1397)(20.756,0.000){4}{\usebox{\plotpoint}}
\put(2144,1397){\usebox{\plotpoint}}
\put(220,200){\usebox{\plotpoint}}
\multiput(220,200)(20.723,1.157){10}{\usebox{\plotpoint}}
\multiput(417,211)(20.729,1.058){9}{\usebox{\plotpoint}}
\multiput(613,221)(20.710,1.367){10}{\usebox{\plotpoint}}
\multiput(810,234)(20.687,1.689){9}{\usebox{\plotpoint}}
\multiput(1006,250)(20.649,2.096){10}{\usebox{\plotpoint}}
\multiput(1203,270)(20.417,3.731){10}{\usebox{\plotpoint}}
\multiput(1400,306)(20.251,4.546){9}{\usebox{\plotpoint}}
\multiput(1596,350)(19.749,6.384){5}{\usebox{\plotpoint}}
\multiput(1695,382)(18.789,8.819){6}{\usebox{\plotpoint}}
\multiput(1793,428)(17.459,11.224){5}{\usebox{\plotpoint}}
\multiput(1891,491)(15.287,14.039){7}{\usebox{\plotpoint}}
\multiput(1989,581)(11.594,17.215){8}{\usebox{\plotpoint}}
\multiput(2088,728)(8.915,18.744){4}{\usebox{\plotpoint}}
\multiput(2127,810)(4.564,20.247){9}{\usebox{\plotpoint}}
\multiput(2166,983)(3.994,20.368){5}{\usebox{\plotpoint}}
\put(2186,1085){\usebox{\plotpoint}}
\sbox{\plotpoint}{\rule[-0.400pt]{0.800pt}{0.800pt}}%
\put(2056,1352){\makebox(0,0)[r]{with NLF}}
\put(2078.0,1352.0){\rule[-0.400pt]{15.899pt}{0.800pt}}
\put(220,189){\usebox{\plotpoint}}
\multiput(220.00,190.40)(10.896,0.514){13}{\rule{15.960pt}{0.124pt}}
\multiput(220.00,187.34)(163.874,10.000){2}{\rule{7.980pt}{0.800pt}}
\multiput(417.00,200.40)(14.226,0.520){9}{\rule{19.800pt}{0.125pt}}
\multiput(417.00,197.34)(154.904,8.000){2}{\rule{9.900pt}{0.800pt}}
\multiput(613.00,208.41)(8.853,0.511){17}{\rule{13.333pt}{0.123pt}}
\multiput(613.00,205.34)(169.326,12.000){2}{\rule{6.667pt}{0.800pt}}
\multiput(810.00,220.41)(8.061,0.509){19}{\rule{12.262pt}{0.123pt}}
\multiput(810.00,217.34)(170.551,13.000){2}{\rule{6.131pt}{0.800pt}}
\multiput(1006.00,233.41)(5.708,0.506){29}{\rule{8.956pt}{0.122pt}}
\multiput(1006.00,230.34)(178.412,18.000){2}{\rule{4.478pt}{0.800pt}}
\multiput(1203.00,251.41)(3.600,0.504){49}{\rule{5.829pt}{0.121pt}}
\multiput(1203.00,248.34)(184.903,28.000){2}{\rule{2.914pt}{0.800pt}}
\multiput(1400.00,279.41)(2.618,0.503){69}{\rule{4.326pt}{0.121pt}}
\multiput(1400.00,276.34)(187.021,38.000){2}{\rule{2.163pt}{0.800pt}}
\multiput(1596.00,317.41)(1.944,0.504){45}{\rule{3.246pt}{0.121pt}}
\multiput(1596.00,314.34)(92.262,26.000){2}{\rule{1.623pt}{0.800pt}}
\multiput(1695.00,343.41)(1.460,0.503){61}{\rule{2.506pt}{0.121pt}}
\multiput(1695.00,340.34)(92.799,34.000){2}{\rule{1.253pt}{0.800pt}}
\multiput(1793.00,377.41)(0.878,0.502){105}{\rule{1.600pt}{0.121pt}}
\multiput(1793.00,374.34)(94.679,56.000){2}{\rule{0.800pt}{0.800pt}}
\multiput(1891.00,433.41)(0.654,0.501){143}{\rule{1.245pt}{0.121pt}}
\multiput(1891.00,430.34)(95.415,75.000){2}{\rule{0.623pt}{0.800pt}}
\multiput(1990.41,507.00)(0.501,0.616){191}{\rule{0.121pt}{1.186pt}}
\multiput(1987.34,507.00)(99.000,119.539){2}{\rule{0.800pt}{0.593pt}}
\multiput(2089.41,629.00)(0.503,1.047){71}{\rule{0.121pt}{1.862pt}}
\multiput(2086.34,629.00)(39.000,77.136){2}{\rule{0.800pt}{0.931pt}}
\multiput(2128.41,710.00)(0.503,1.583){71}{\rule{0.121pt}{2.703pt}}
\multiput(2125.34,710.00)(39.000,116.391){2}{\rule{0.800pt}{1.351pt}}
\multiput(2167.41,832.00)(0.505,2.111){33}{\rule{0.122pt}{3.480pt}}
\multiput(2164.34,832.00)(20.000,74.777){2}{\rule{0.800pt}{1.740pt}}
\end{picture}
%\end{document}
%\newpage
%\input{fig1.tex}
%\begin{center}
%Fig.
%\end{center}

\begin {thebibliography}{99}
   
\bibitem{boy}
Boyle, B.J., Shanks, T., and  Peterson,B.A.,  \emph{M.N.R.A.S}, {\bf 235}, 935 (1988) 
%
\bibitem{bri}
Brimer, T.G. and  Sanders, R.H., \emph {A$\&$A}, {\bf 274}, 96 (1993)
%
\bibitem{fuku0}
Fukugita, M., Futamse, T.  and  Kasai, M.,  \emph{M.N.R.A.S}, {\bf 246}, 24p (1990)
%
\bibitem{fuku1}
Fukugita, M. and  Turner, E.L., \emph{M.N.R.A.S}, {\bf 253}, 99 (1991)
%
\bibitem{fuku2}
Fukugita, M.,  Futamse, T.,  Kasai, M., and Turner, E.L., \emph{Ap.J}, {\bf 393}, 3 (1992) 
\bibitem{koch0}
Kochanek, C.S., \emph {Ap.J}, {\bf 419}, 12 (1993)
%
\bibitem{koch1}
Kochanek, C.S., \emph {Ap.J}, {\bf 436}, 56 (1994)
%
\bibitem{koch2}
Kochanek, C.S., \emph{Ap.J}, {\bf 466}, 638 (1996)
%
\bibitem{krau}
Krauss, L.M. and Turner, M.S., 1995, Preprint astro--ph/{9504003} 
%
\bibitem{naka}
Nakamura, T.T and  Suto, Y., UTAP Preprint No. {236/96}.
%
\bibitem{press0}
Press, W.H., and  Gunn, J.E.  \emph{Ap.J}, {\bf 185}, 397 (1973)
%
\bibitem{press1}
Press, W.H. and Schechter, P.,  \emph{Ap.J}, {\bf 187} , 487 (1974)
%
\bibitem{refs}
Refsdal, S.,  \emph{M.N.R.A.S}, {\bf 128},  295 (1964)
%
\bibitem{rocca}
Rocca-Volmerange, B., and  Guiderdoni, B.,  \emph{M.N.R.A.S}, {\bf 247}, 166 (1990)
%
\bibitem{turn0}
Turner, E.L., Ostriker, J.P. and Gott, J.R.,III., \emph{Ap.J}, {\bf 284}, 1 (1984) (TOG).
%
\bibitem{turn1}
Turner, E.L.,  \emph{Ap.J}, {\bf 365}, L43 (1990)













   \end {thebibliography}
\end {document}
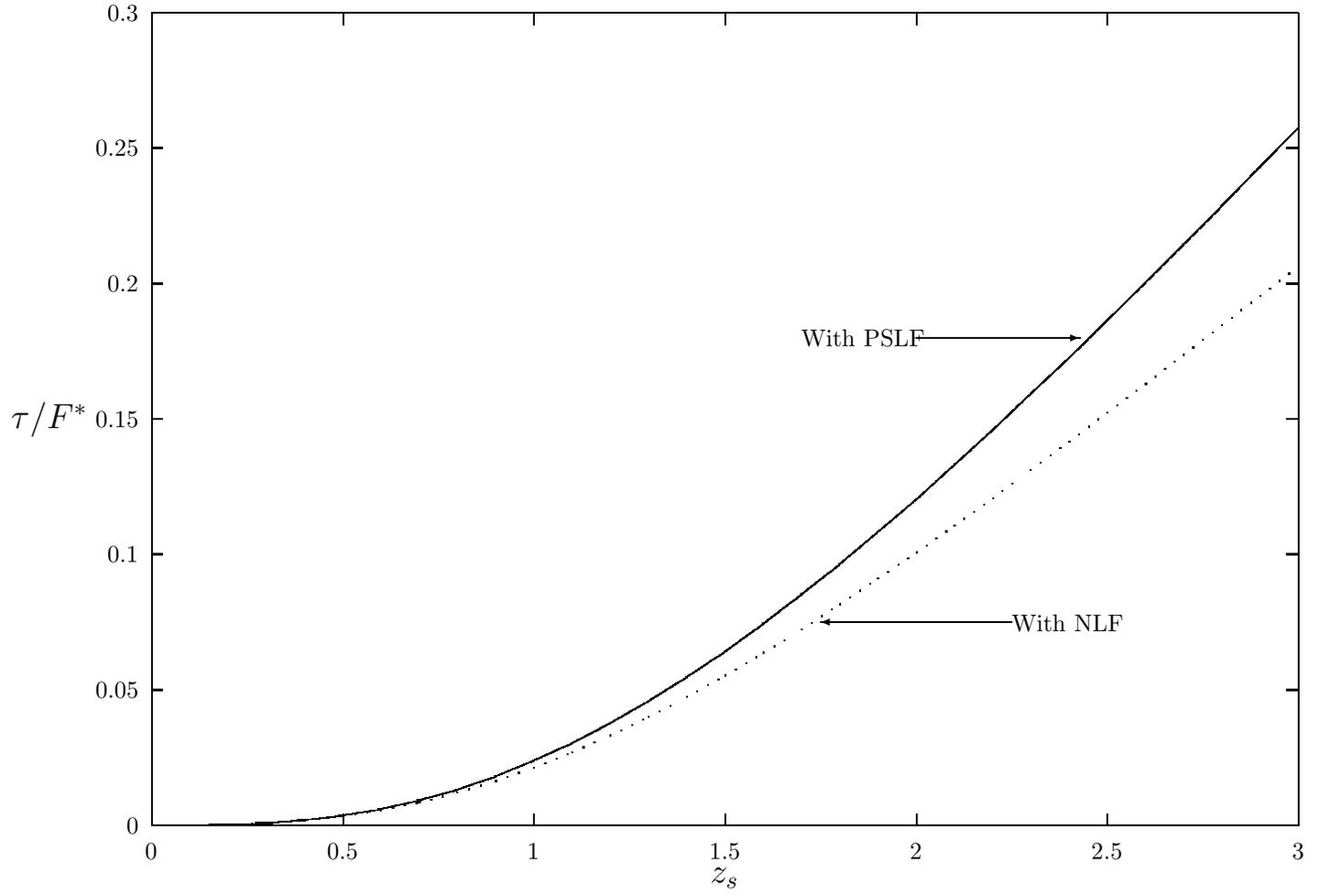
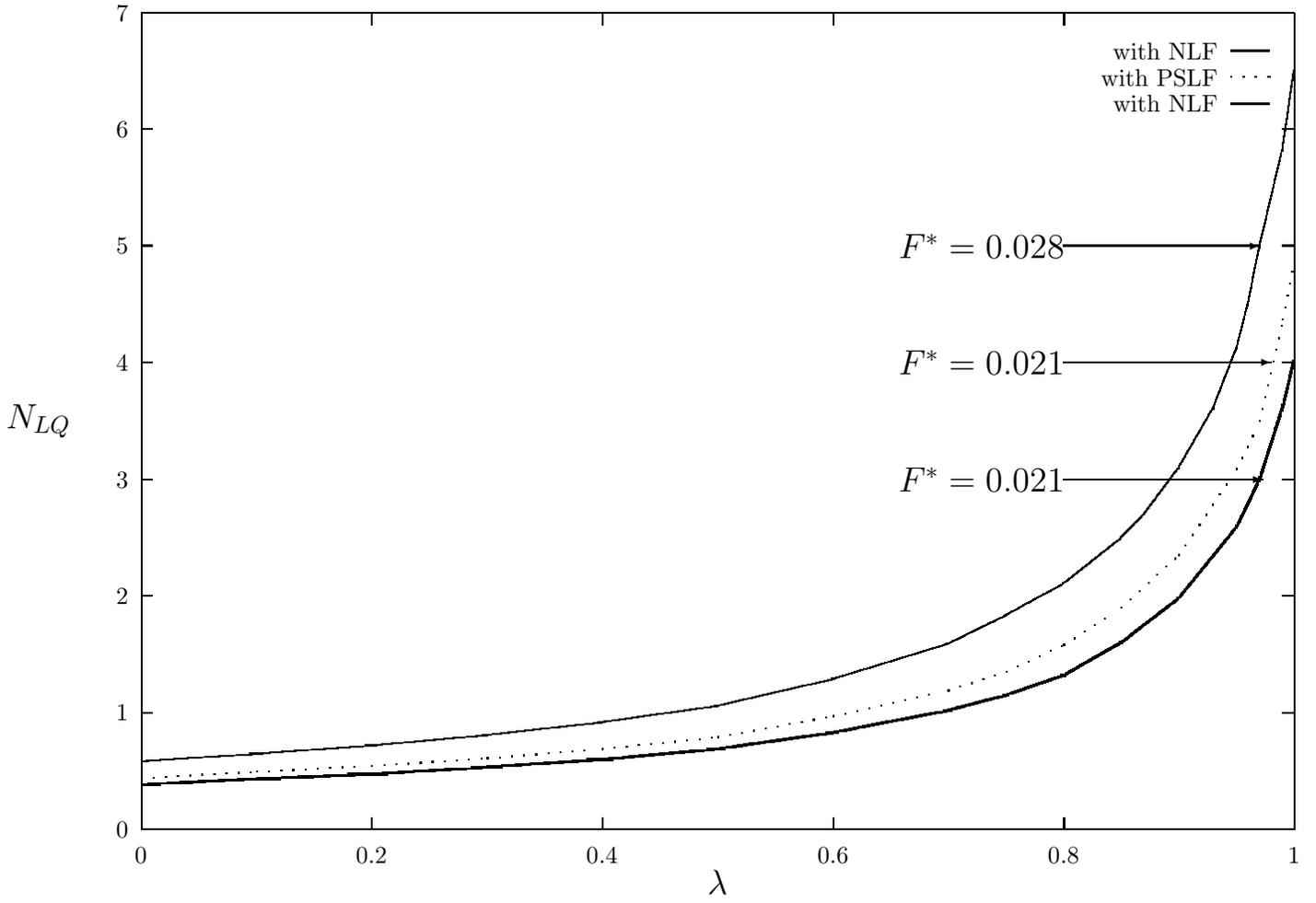
